\documentclass[twocolumn,showpacs,preprintnumbers,amsmath,amssymb]{revtex4}
\usepackage{graphicx}
\usepackage{dcolumn}
\usepackage{bm}

\begin{document}

\title{A definitive number of atoms on demand: controlling the number of atoms in a-few-atom magneto-optical trap}
\author{Seokchan Yoon, Youngwoon Choi, Sangbum Park, Jaisoon Kim, Jai-Hyung Lee} 
\author{Kyungwon An}
\email{kwan@phya.snu.ac.kr}
\affiliation{School of Physics, Seoul National University, Seoul 151-742, Korea}
\date{\today}

\begin{abstract}
A few $^{85}$Rb atoms were trapped in a micron-size magneto-optical trap 
with a high quadrupole magnetic-field gradient and the number of atoms was precisely controlled by suppressing stochastic loading and loss events via real-time feedback on the magnetic field gradient. The measured occupation probability of single atom was as high as 99 \%. Atoms up to five were also trapped with high occupation probabilities.
The present technique could be used to make a deterministic atom source.
\end{abstract}
\pacs{Valid PACS appear here}
\maketitle

In recent years there has been a strong interest in trapping and manipulating single neutral atoms localized spatially with a small kinetic energy. 
In addition, a deterministic atom source for single or an exactly known number of atoms is necessary for various experiments in quantum optics, cold collision, precision measurements and quantum information \cite{quantum_info}. 
Several groups have demonstrated trapping single atoms and controlling position of atoms by means of magneto-optical trap (MOT) \cite{singleatomMOT_Kimble,singleatomMOT_Meschede,singleatomMOT_Ertmer} or optical dipole trap \cite{conveyor}. 
Reduction of the probability of trapping more than one atom as a result of intra-trap two-body collision effects was also observed in a tight-focused dipole trap superimposed on a MOT \cite{superimposed_dipoletrap}. 

Because of the stochastic nature of loading and loss events in MOTs, all of the above works rely on ultrahigh vacuum for a relatively long trap lifetime and a low loading rate for suppression of random multi-atom loading events. As a result, one-atom trap events are rare, and thus  very long waiting times are needed in order to capture single atoms. Furthermore, trapping more than one atom deterministically for an extended period of time was not possible in those experiments.
Recently, deterministic production of single chromium atoms in a MOT with a controlled feedback on loading and loss rates has been reported \cite{beamfeedback}. In that report, a chromium atomic beam from an evaporator was manipulated for controlling the loading rate of the trap. 

In the present work, we have realized a deterministic source of rubidium atoms based on a micron-sized MOT in a vapor cell with high occupation probabilities of single or any desired number atoms, up to five atoms, controlled by a magnetic-field-gradient feedback technique.
In a MOT with a high magnetic field gradient, as the field gradient is increased, both the capture velocity $v_{\rm cap}$ and the trapping area $A_{tr}$ of the trap decrease, and thus the loading rate $R_{L}$, proportional to $A_{tr}v_{\rm cap}^4$ in one-dimensional model \cite{strongmagneticfield}, decreases rapidly, enabling great reduction of the average number of trapped atoms.
Strong magnetic field gradient also plays an important roles in making a well localized atomic source since the trap size is inversely proportional to the square root of magnetic field gradient.
The high magnetic field gradient is, however, not the only requirement for trapping a well-defined small number of atoms in a MOT since the number of atoms in the trap still fluctuates randomly by stochastic loading and loss processes even under the high magnetic-field gradient. 
We have reduced the random fluctuations in the number of atoms by controlling the loading rate with real-time feedback on the magnetic field gradient.

\begin{figure}[b]
\includegraphics[width=3.3in]{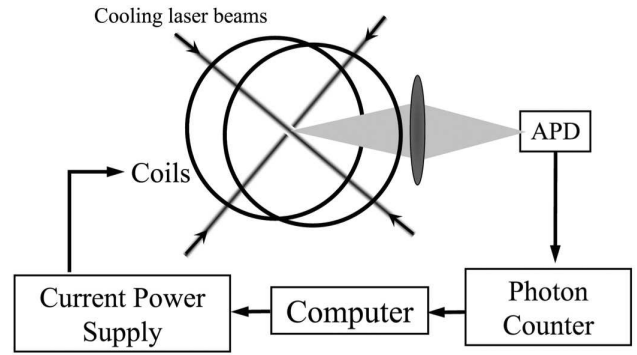}
\caption{Schematic of the magnetic-field gradient feedback for a few-atom MOT.}
\label{fig:figure1}
\end{figure}

A schematic of our experiment is shown in Fig.~\ref{fig:figure1}. 
Our MOT follows a standard MOT configuration with six cooling laser beams and a repump laser.
The total power of cooling laser beams was set to 1.2 mW with a beam size of 3 mm in diameter and the frequency of cooling laser was tuned to -6 MHz from the D2 transition of $^{85}$Rb.
The base pressure of a ultrahigh vacuum chamber containing the trap was maintained in the range of 10$^{-11}$ Torr. 
A current-driven Rb getter dispenser was used to produce background Rb atoms to be trapped.
Two water-cooled electromagnets in anti-Helmholtz configuration were employed in order to produce a high magnetic-field gradient up to 600 G/cm.

The exact number of atoms in the trap was measured from the fluorescence light emitted by the MOT. 
The fluorescence light was collected with an objective lens with a numerical aperture of 0.23, mounted inside the vacuum chamber. 
After passing through a focusing lens and an optical bandpass filter centered at 780 nm with a 10-nm bandwidth, the collected fluorescence was detected by an avalanche photodiode operating in photon-counting mode with a time bin of 100 ms. 
The overall photon collection efficiency including optical losses at various stages in our detection system was about 0.9\%.
Fluorescence images of trapped atoms were also taken by a digital charge-coupled-device camera. When the magnetic field-gradient is higher than 400 G/cm, the measured trap size was 5 $\mu$m (FWHM), which is mostly determined by the optical resolution of our imaging optics.
\begin{figure}[t]
\includegraphics[width=3.3in]{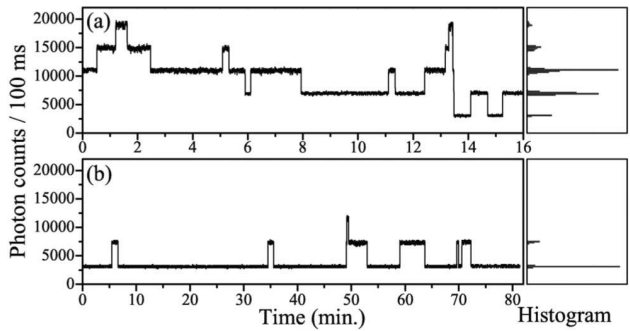}
\caption{Typical time traces of fluorescence signal from a few-atom MOT. The magnetic-field gradient is 210 G/cm in (a) and 420 G/cm in (b) whereas the average number of atoms is 1.2 in (a) and 0.18 in (b). The current applied to the Rb dispenser was fixed at 1.9 Amps during the measurements.}
\label{fig:figure2}
\end{figure}

Figures \ref{fig:figure2} (a) and (b) show partial time traces of typical step-wise fluorescence signals from the MOT, obtained by the avalanche photo diode for two different loading rates set by different magnetic-field gradients for the same Rb background pressure. Such step-wise signals are characteristics of a few-atom MOT with the fluorescence levels quantized according to the instantaneous number of atoms in the trap.
We can measure both the loading rate $R_L$ and the one atom loss rate $\Gamma_1$ with background gas by counting the number of one-step-upward or one-step-downward transitions in the fluorescence signal since such transitions correspond to addition or subtraction of one atom in the trap, respectively. 
The loading rates measured in this way for complete time traces corresponding to Figs.\ \ref{fig:figure2}(a) and (b) are $1.1\times 10^{-2} s^{-1}$ and $1.6\times 10^{-3} s^{-1}$, respectively. The one atom loss rate is $0.9\times 10^{-2} s^{-1}$.

A lifetime $T_1$ of single-atom trap is defined as the average time during which a single atom remains in the trap. 
Because both loading and loss events change the number of trapped atoms, $T_1=[R_L+\Gamma_1]^{-1}$.
Although it is desirable for a given $\Gamma_1$ to increase $T_1$ as much as possible by decreasing $R_L$, $R_L$ cannot be made too small since it then takes too long to reload the empty trap. The waiting time for loading single atom $T$ is given by $T=R_L^{-1}$.
The measured trap lifetime and waiting time for one atom in Fig.\ \ref{fig:figure2}(b) were 110 s and 590 s, respectively.

\begin{figure}[t]
\includegraphics[width=3.3in]{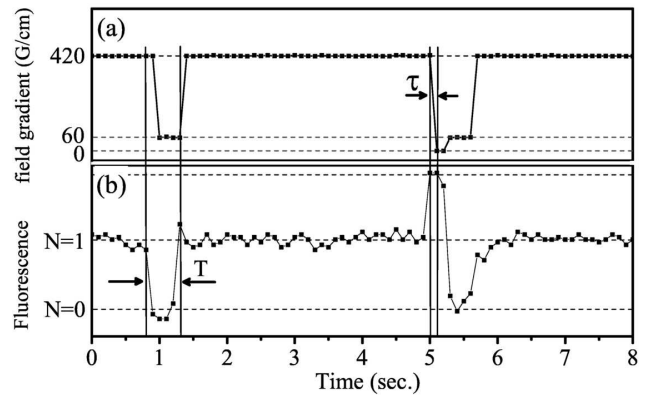}
\caption{Feedback in action. (a) Time trace of the magnetic field gradient and (b) the fluorescence signal from trapped atoms. 
The field gradient can be changed with a rise/fall time $\tau$ of 100 ms. The waiting time T for one atom was about 600 ms.}
\label{fig:figure3}
\end{figure}

For the purpose of active controlling the number of atoms, the loading rate is controlled by changing the magnetic-field gradient. The sequence of field-gradient feedback is depicted in Fig.\ \ref{fig:figure3}.
At first, atoms are loaded into the trap at a low magnetic-field gradient of 60 G/cm with a short waiting time {\em i.e.}, with a relatively high loading rate. 
The number of atoms measured via the fluorescence signal is used as a trigger signal for changing the field gradient. After a given number of atoms are loaded in the trap, the field gradient is rapidly increased to a high value of 420 G/cm in order to reduce the loading rate almost to zero. 
At this high field gradient, practically no atoms can be loaded into the trap and only loss events occur.
As soon as any atom escapes from the trap and thus the fluorescence falls below a preset level corresponding to the given number of atoms, the field gradient is rapidly decreased to a low value at which atoms can be loaded into the trap quickly. 
After waiting for the desired number of  atoms to be captured in the trap, the field gradient is set back to the high value again in order to shut off the loading rate.
If the number of trapped atoms are more than the preset number, the magnetic field is completely turned off for a short time ($\sim$100 ms) in order to flush out all of the trapped atoms and then turned on again for reloading atoms again. 

\begin{figure*}[t]
\includegraphics [width=6.5in]{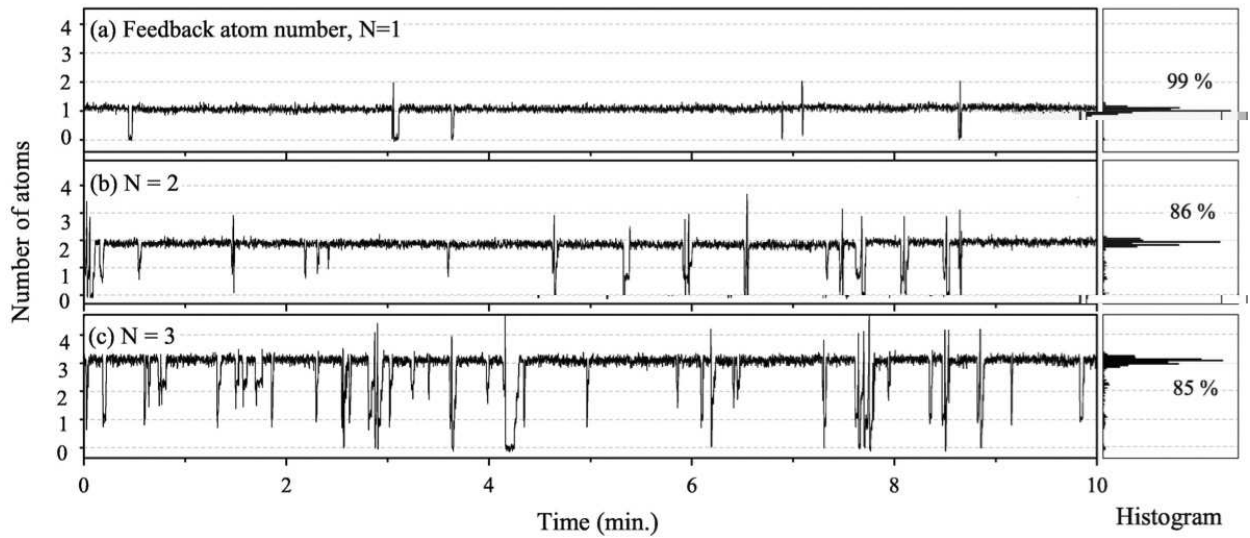}
\caption{Results of the magnetic field gradient feedback. The atom number is fixed to one in (a), two in (b) and three in (c). Each histogram shows the occupation probability for the preset feedback atom number. Photon count rate corresponding to one atom is about 17,000 cps.}
\label{fig:figure4}
\end{figure*}

At a fixed Rb background vapor pressure with a constant current of 2.1 A applied to the Rb dispenser, the loading rate rapidly decreases to 0.001 s$^{-1}$ as the field gradient increases to 420 G/cm, resulting in a waiting time of about 17 minutes. In contrast, when the field gradient is 60 G/cm, one atom can be loaded in a waiting time of only 600 ms.

The time traces of fluorescence signal when the feedback is employed are shown in Fig.~\ref{fig:figure4} with their level histograms for various preset feedback atom numbers.
The observed occupation probabilities for a single atom was as high as 99 \%.
The trap lifetime for single atom was about 99 s, limited only by the collisions with background gas molecules, and thus $\Gamma_1=1.0\times 10^{-2} s^{-1}$.
However, the occupation probabilities (and trap lifetimes) of two and three atoms were measured to be 86 \% (29 s) and 85 \% (14 s), respectively.
In addition, the maximum occupation probabilities achieved by the feedback for four and five atoms were reduced to 77 \% and 76 \%, respectively.
These trap lifetimes are much shorter than $(\Gamma_1 N)^{-1}$ with $N \geq 2$ due to light-induced intra-trap two-atom collisional losses \cite{two_atom_collision}.
If we denote the two-atom collision rate as $\Gamma_2$, the total loss rate for $N$ atoms is given by $\Gamma_1 N +\Gamma_2 N(N-1)/2$.
From the above trap lifetimes for $N=2,3$, we then obtain $\Gamma_2\simeq 1.4\times 10^{-2} s^{-1}$.
The intratrap two-atom collision rate $\Gamma_2$ depends on the intensities and detunings of cooling and repumping lasers \cite{minimize_collision}. 
The $\Gamma_2$ could be minimized by an order of magnitude by simply adjusting the laser intensities.

Under the present experimental conditions, we could control the atom number up to five atoms with high occupation probabilities. However, our field gradient feedback itself is not limited to five atoms. More than five atoms can be prepared with high efficiencies by enhancing the imaging accuracy at the low field gradient and by improving the unloading algorithm for the case of overloaded atoms \cite{improving}.

Our trap can be used as a deterministic atom source for many applications such as quantum information and nanotechnology. Other applications include studying coherent collective effects of precisely known small number of atoms and quantitative analysis of two-body cold collision dynamics in a micron size trap \cite{two-body-collision}.

This work was supported by Korea Research Foundation Grant (KRF-2005-070-C00058), KOSEF Grant (NRL-2005-01371) and grant from the Ministry of Science and Technology of Korea.

\end{document}